# Silicon metasurfaces for third harmonic geometric phase manipulation and multiplexed holography

Bernhard Reineke[1], Basudeb Sain[1], Ruizhe Zhao[2], Luca Carletti[3], Bingyi Liu[4], Lingling Huang[2], Costantino De Angelis[5], Thomas Zentgraf[1]

*1 Department of Physics, Paderborn University,*
*Warburger Straße 100, D-33098 Paderborn, Germany*

*2 School of Optics and Photonics, Beijing Institute of Technology, Beijing 100081, China*

*3 Dept. of Information Engineering, University of Padova, 35131 Padova, Italy*

*4 Institute of Modern Optics, Department of Physics, Harbin Institute of Technology, Harbin 150001, China*

*5 Dept. of Information Engineering, University of Brescia, 25123 Brescia, Italy*
*National Institute of Optics (INO), CNR, 25123 Brescia, Italy*

*Corresponding author: Thomas Zentgraf, thomas.zentgraf@uni-paderborn.de*

**Abstract**

Nonlinear wavefront control is a crucial requirement in realizing nonlinear optical applications with metasurfaces. Numerous aspects of nonlinear frequency conversion and wavefront control have been demonstrated for plasmonic metasurfaces. However, several disadvantages limit their applicability in nonlinear nanophotonics, including high dissipative loss and low optical damage threshold. In contrast, it has been shown that metasurfaces made of high-index dielectrics can provide strong nonlinear responses. Regardless of the recent progress in nonlinear optical processes using all-dielectric nanostructures and metasurfaces, much less advancement has been made in realizing a full wavefront control directly with the generation process. Here, we demonstrate the nonlinear wavefront control for the third-harmonic generation with a silicon metasurface. We use a Pancharatnam-Berry phase approach to encode phase gradients and holographic images on nanostructured silicon metasurfaces. We experimentally demonstrate the polarization-dependent wavefront control and the reconstruction of an encoded hologram at the third-harmonic wavelength with high fidelity. Further, we show that holographic multiplexing is possible by utilizing the polarization states of the third harmonic generation. Our approach eases design and fabrication processes and paves the way to an easy to use toolbox for nonlinear optical wavefront control with all-dielectric metasurfaces.

**Introduction**

Metasurfaces enable the potential of efficient generation of new frequencies and simultaneous wavefront control of the generated light. Unfettered from the cumbersome phase-matching limitations and featuring a unique control over nonlinear fields, plasmonic metasurfaces have been extensively used to control nonlinear optical processes at the nanoscale, such as harmonic generation[1–3], wave mixing[4], nonlinear imaging[5], nonlinear beam shaping[6,7], and holography[8], during the past decade[1,9]. Plasmonic metasurfaces, which are mostly made of metallic elements, allow a large degree of control of the optical nonlinearity by engineering subwavelength meta-atoms with precise polarization control and strong local field enhancement. However, high dissipative losses and a low optical damage threshold severely restrict the obtainable nonlinear conversion efficiency of nonlinear optical processes. Recently, high refractive index all-dielectric metasurfaces have emerged as a new class of ultrathin components for nonlinear optical applications. They can overcome the aforementioned limitations by having lower dissipative losses, the ability to withstand much higher pump field intensities, and even can show larger nonlinear coefficients. It has been shown that all-dielectric resonators and metasurfaces can efficiently confine light into subwavelength volumes [10,11] and support even stronger nonlinear optical responses enabled by strong resonances along with inherent high material nonlinearities [12–18]. In this context, all-dielectric nanophotonics has been used to demonstrate many interesting phenomena like efficient optical frequency conversion and mixing, vector beam generation, and shaping the radiation pattern by exploiting Kerker condition or using different crystallographic axis [18–22].

Regardless of the large progress in nonlinear optics with plasmonic metasurfaces and nanostructures[1,23] much less advancement has been made in wavefront shaping by nonlinear all-dielectric metasurfaces. In contrast to plasmonic nano-antennas, dielectric nano-resonators possess larger volumes which lead to longer propagation lengths through the structures as well as higher packing densities which increase coupling effects. Therefore, phase matching effects and crosstalk between the localized modes in the structures become nonnegligible. Furthermore, the field in each nano-resonator is a superposition of various resonances with different far-field radiation patterns. Therefore, controlling the phases and the direction of the generated light, required for the nonlinear wavefront control, become a complex problem, since the interaction with the neighboring nano-resonators has to be taken into account. Very recently, Wang et al. and Gao et. al. demonstrated the wavefront control of the third-harmonic light generated from all-dielectric silicon metasurfaces [24,25]. Their approach relies on the generalized Huygens' principle in nonlinear optics. Utilizing the resonant behavior of a single silicon nano-post, they have been able to design a set of individual structures that result in different phase delays for the locally generated third harmonic signal. However, to smoothly cover the full phase range of 0 to 2π, a large set of structures has to be designed. Such a design is a computationally heavy task since a single nano-post possesses several geometric degrees of freedom and the optical properties for the nonlinear process need to be matched with the desired phase and amplitude. Therefore, the demonstrated method of nonlinear phase control depends sensitively on the precise geometry and refractive index of the nanostructures, which also makes the fabrication challenging and the optical properties susceptible to fabrication tolerances.

In contrast to the nonlinear phase control by the Huygens' principle, the concept of a geometric phase metasurface, also known as Pancharatnam-Berry (PB) phase metasurface, can achieve a continuous phase change by using the degree of rotation of an optimized design of a nano-post, which can in many ways simplify both the design and fabrication processes of the desired metasurface[9]. The working principle of a nonlinear PB-phase metasurface has already been shown in several plasmonic systems associated with applications like nonlinear optical holography [8] or beam profile manipulation [5,26]. The geometric phase approach to control the phase of the generated second harmonic generation (SHG) light from an all-dielectric silicon metasurface was revealed by Bar-David et al. [27] The SHG signal was mostly generated from the surface of the nano-resonators and, hence, does not utilize the field enhancement inside the nano-resonator's volume. In addition, silicon has an inversion symmetry and does not support SHG within the dipole approximation from the bulk material. Therefore, the question appears if a nonlinear all-dielectric PB-phase metasurface that utilizes the large third-order nonlinear coefficients of silicon and the local field enhancement inside the nano-resonator is feasible, but circumventing the contrary effects from phase matching and crosstalk. In this work, we demonstrate the phase-controlled generation of third-harmonic light from an all-dielectric amorphous silicon metasurface made of nanofins with a twofold rotational (C2) symmetry. Similar to plasmonic metasurfaces, we show that by using different orientation angles of the silicon nanofins it is feasible to encode a defined spatial phase gradient into the generated third-harmonic light despite the larger structure thickness and the appearing phase matching issues. We further extend the principle to demonstrate a nonlinear holographic metasurface, showing the reconstruction of holographic images by the third harmonic generation (THG) signal. Our approach to utilize the PB-phase effect for circularly polarized light together with the bulk nonlinearity allows simplification in the design of the metasurfaces and provides higher tolerances for the fabrication.

**Design of the metasurface**

In the nonlinear regime, the PB-phase approach connects the rotation angle of a single nano-resonator to the phase shift at the generated harmonic light under coherent circularly polarized illumination. From the plasmonic counterpart, it is known that a rotation angle $\phi$ of a plasmonic antenna can induce an abrupt phase shift θ in the *n*-th harmonic signal [28]. The general relationship between θ, *n*, and $\phi$ is given by $θ = \sigma(n \pm 1)\phi$. According to this relation, a phase shift can be introduced at the co- or cross-circularly polarised generated light (identified by the $\pm$ sign, respectively) compared to that of the fundamental wave. In the case of THG ($n = 3$) and a nanofin with C2 rotational symmetry, we expect the phase factors $θ_{co} = 2\sigma\phi$ and $θ_{cross} = 4\sigma\phi$ in the co- and cross-polarized states, depending on the rotation angle, respectively.

Figure 1a summarizes the principle of the phase shift of the generated light with antenna's rotation. Third harmonic light with both left and right circular polarizations (LCP, RCP) is generated when a right circularly polarized (RCP) light beam is incident on a single resonator. However, the co-polarization (RCP-RCP) carries different phase information than the cross-polarization (RCP-LCP).

On the basis of the aforementioned discussion, we designed and fabricated an amorphous silicon metasurface, made of nanofins with C2 symmetry, to be operated at the fundamental wavelength 1240 nm. We encoded a phase gradient of 0 to $2\pi$ within an array of length 4.6 µm in the co-polarization state whereas, a phase gradient of 0 to $4\pi$ was encoded in the cross-polarization state, compared to the fundamental wave. The induced phase shifts for the different combinations of the input and output polarizations lead to differently polarised diffraction spots at different angles; 5.15° ($2\pi$, co-polarization) and 10.30° ($4\pi$, cross-polarization) for the TH wavelength, schematically shown in

Figure 1b. As a reference, we also fabricated a metasurface without a phase grating, where every antenna has the same orientation and, therefore, the THG constructively interferes at an angle of 0° for both polarization combinations. (for details see Supplementary Material).

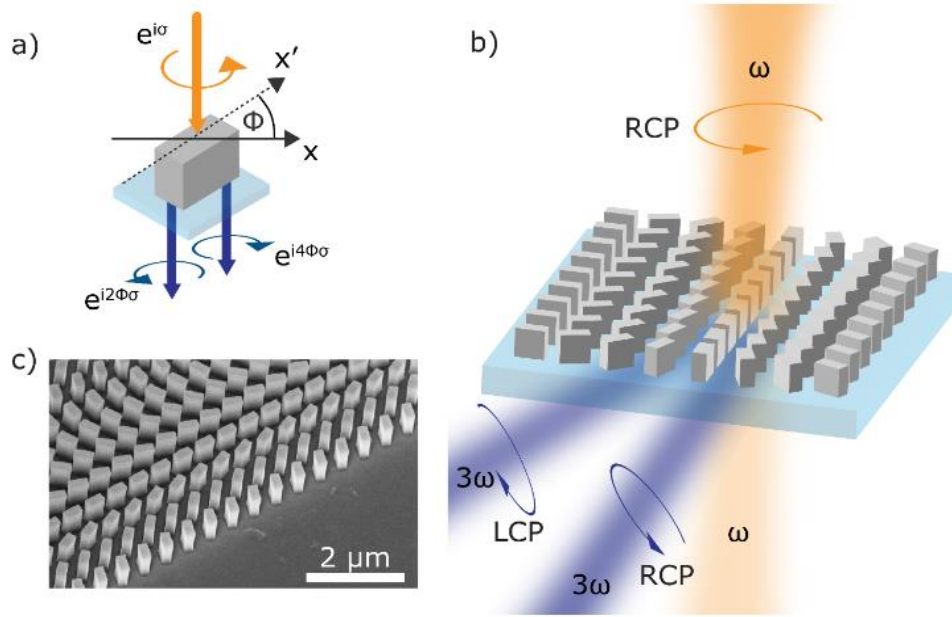

**Figure 1. a)** Schematic illustration of the generation of third harmonic light with antenna's rotation. Upon rotation the generated nonlinear signal gains phases of $\theta_{co} = 2\sigma\phi$ and $\theta_{cross} = 4\sigma\phi$ in co- and cross-polarization states, respectively. **b)** Schematic of the working principle of a geometric-phase silicon metasurface. A fundamental beam of right circularly polarised (RCP) light is incident onto the metasurface. The nanostructures within the metasurface can generate a third harmonic light. Further, they are arranged onto the surface to encode a phase gradient of 0 to $2\pi$ (0 to $4\pi$) at the RCP (LCP) generated third harmonic light. **c)** Scanning electron microscopy image of a part of the fabricated Si metasurface.

The fabricated metasurface with the phase gradient consists of identical Si nanofins, with a size of 400x200x650 nm³ (LxWxH) on top of an ITO coated glass substrate. The nanofins were arranged in a 100x100 µm² square array with a period of 575 nm whereas the spatial orientation of each nanofin varies (Figure 1c). More fabrication details can be found in the Supplementary Material.

Furthermore, we designed and fabricated two additional samples that have encoded the phase distribution of a computer-generated hologram (CGHs), namely the letter "X" and phase distribution of a multiplexed hologram. The use of the PB-phase provides two distinct nonlinear phases in the THG signal for the two generated polarization states. Hence, it allows encoding two different images into a single metasurface. For our demonstration, we use the schematic image of a 'sun' and a 'cloud', which can be read out by switching between co- and cross-polarization states. Multiplexed holography exploits the property of our nanofin design, that co- and cross-polarization states are orthogonal and carry different phase information. Details about the holographic multiplexing can be found in the Supplementary Material.

The holographic metasurfaces use the same nanofin geometry as for the phase gradient but the spatial rotation of each nanofin is now determined by the phase distribution of the CGH. The phase distributions were calculated using a modified Gerchberg-Saxton algorithm (for details see the Supplementary Material). To simplify the measurements, the holograms were designed to reconstruct in the in Fourier space. This allows us to unify the measurement setup for the THG phase gradient metasurface and the THG holographic metasurfaces.

**Experimental results**

For the experimental investigation of the nonlinear properties of our samples, we used the experimental set-up schematically drawn in Figure 2a. The sample was illuminated by a slightly focused circularly polarised laser beam at wavelengths between 1200 nm and 1350 nm (fundamental wavelength). As a laser source, we used an optical parametric oscillator (OPO) with a typical pulse length of 200 fs and a repetition rate of 80 MHz synchronously pumped by a TiSa femtosecond laser. The THG signal of the metasurface was collected by an infinity-corrected microscope objective lens with NA=0.6. By imaging the back focal plane of the microscope objective onto a CMOS camera and filtering the polarization state of the THG light, we measured the diffraction angle of the frequency-converted light. The diffraction angle $\alpha_{\text{diff}}$ of the THG signal at the camera location is given by:

$$\alpha_{\text{diff}} = \text{atan}\left(\frac{\Delta x}{W_{\text{Obj}}}\right) \qquad 1$$

whereas $\Delta x$ is the lateral distance between the centre of the observable Fourier space and the observed diffraction spots. $W_{\text{Obj}}$ (here $W_{\text{Obj}} = 4.75 \text{ mm}$) is the working distance between the sample and the objective lens. Since both lenses behind the objective lens have the same focal length no additional magnification correction for the imaging system has to be applied. Note that a short pass filter has been used in the detection beam path to suppress the fundamental beam.

Figure 2b and Figure 2c display the THG measurement results for the different co- and cross-polarization states from the metasurfaces with and without phase gradient at a wavelength of 1240 nm, respectively. For the sample with the designed phase gradient in the THG signal (Figure 2b), we observe diffraction spots at angles of $\pm(5.36\pm0.01)°$ for the co-polarization states. This value is close to the designed diffraction angles of $\pm5.01°$. In the cross-polarization states, the diffraction spots appear at $\pm(10.10\pm0.01)°$, which is again close to the designed angle $\pm10.3°$. The different diffraction angles for the co- and cross-polarized THG signal is a consequence of the different accumulated PB-phase at the metasurface.

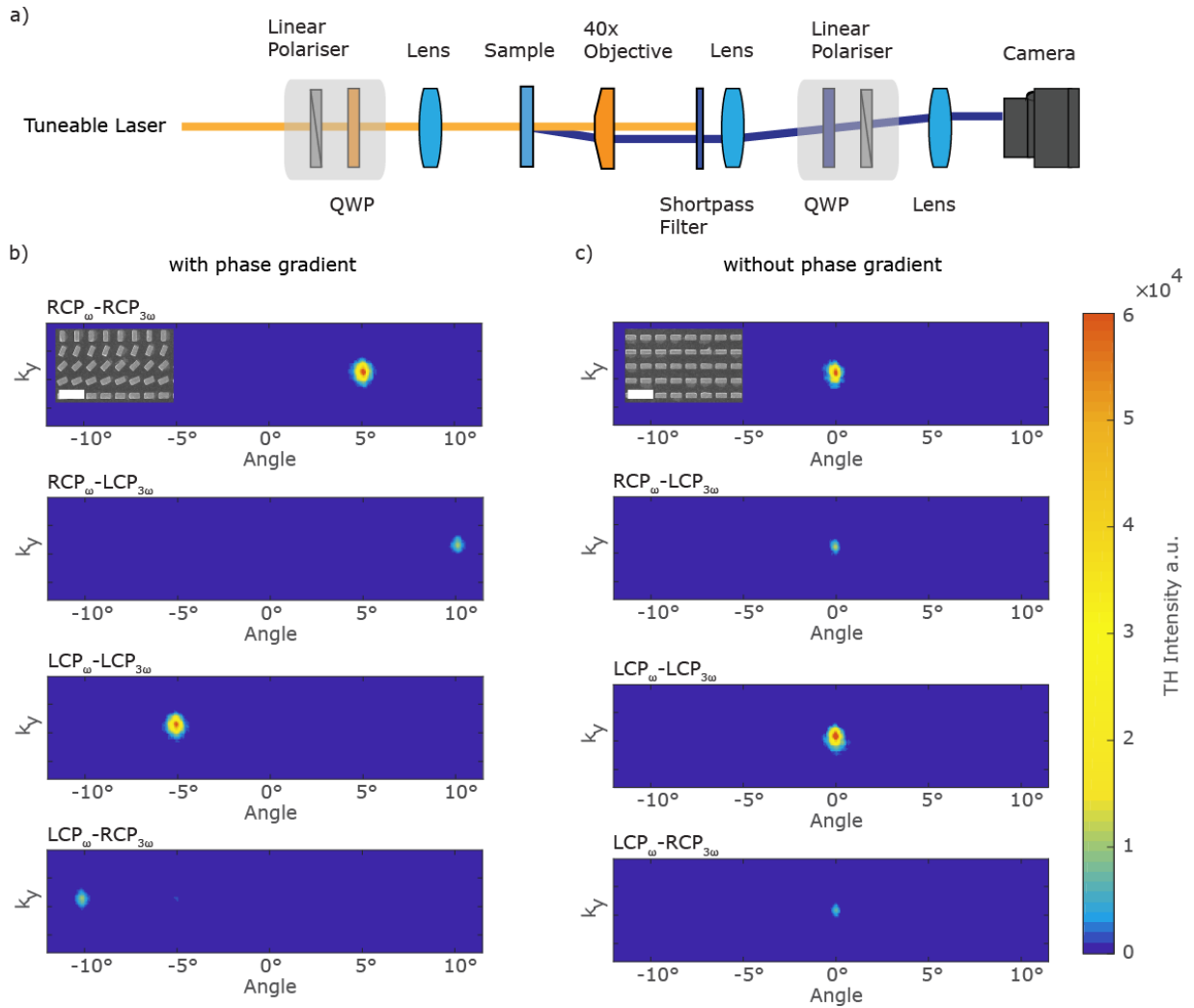

**Figure 2. a)** Schematic illustration of the optical measurement setup. The first combination of a linear polarizer and a quarter-wave plate (QWP) prepares the circularly polarized input light at the fundamental wavelength. The second combination analyses the output THG light. The lens in front of the sample focuses the light onto the sample, while the lenses behind the objective lens image the back focal plane of the objective onto the CMOS camera. The THG light is collected by a 40x/NA0.6 microscope objective. **b)** Third harmonic diffraction spots generated by the phase-gradient metasurface. The THG signals are measured with different combinations of circularly polarized input and output light. The inset shows the top view of the corresponding metasurface. The scale bar represents 1 µm. **c)** Third harmonic diffractions spots for metasurface without phase-gradient for different combinations of circularly polarized input and output light. The inset shows the top view of the metasurface. Note that the color scheme was modified to increase the visibility of the weaker THG diffraction spot in the image.

Furthermore, we observe that the diffraction spots for the co-polarization states of the THG signals are approximately 3 times brighter than in the cross-polarization state. In contrast, a metasurface without phase gradient (Figure 2c), where all nanofins have an identical orientation, provides no additional diffraction orders except the zeroth-order THG signal. Here, the THG signal from all antennas radiate with the same geometric phase and interfere constructively at an angle of 0°.

Figures 3a-c show the obtained results from wavelength-dependent measurements of the THG signal strength from the metasurface samples with phase gradient (rotated nanofins), without phase gradient (all nanofins have the same orientation) and for an unstructured silicon film of the same thickness as the nanofins (650 nm) for the co- and cross-polarization states of the THG signal. For analyzing the wavelength dependence in more detail, we plot the THG intensity of the phase-gradient metasurface sample for each diffraction spot (Figure 3a). If the tailored phase in the THG signal arises from the PB-phase, the observed diffraction of the beam should be observable over a broad wavelength range. Therefore, we analyzed the diffraction spots for a wavelength range between 1200 nm and 1350 nm for the fundamental wavelength. Due to the wavelength change the diffraction angle changes from 5.02° to 5.63° (co-pol.) and 9.97° to 11.11° (cross-pol.) from 1200 nm to 1350 nm, respectively (see Supplementary Material). For simplicity, the plot in Figure 3a shows the integrated THG intensities for the corresponding diffraction angle. We observe that the THG intensity of the co-polarization signal stays at a nearly constant level independent on the used fundamental wavelength. The cross-polarization THG intensity is always lower and reaches a local minimum at around 1275 nm. Similar results are obtained for a metasurface without the phase-gradient for the THG light (Figure 3b). Note that for this metasurface the integrated THG intensity was always obtained from the zeroth-order spot.

Further, we observe that compared to a 650-nm-thick amorphous silicon film, the THG signal of the metasurface is enhanced by a factor of ~40 (Figure 3c) for the co-polarization state. The silicon film itself shows a negligible amount of THG in the co-polarization state, which is close to our detection limit. Note, that the low THG signal from the silicon film is anticipated as amorphous silicon is an isotropic medium. In a perfect isotropic medium, the THG under circularly polarized excitation is forbidden due to the dipole selection rules.[29,30] This leads to the conclusion that the observed THG signal is entirely due to the nanostructured amorphous silicon nanofins since they break the isotropy of the film. In this context, it is worthy to mention that the bare ITO-coated glass substrate, which is used as a substrate for the silicon metasurface, has also a negligible nonlinear conversion efficiency for the same reason (Figure 3c).

The results for the phase-gradient metasurface demonstrate that the spatial phase of the THG can be tailored by the spatial orientation of identical nanofins. Next, we use the same concept to encode more complex spatial phase profile into the metasurface. Here, we use the silicon metasurface as nonlinear holograms for the letter 'X' and for a multiplexed hologram for the schematic image of a sun and a cloud. For the measurement of the holographic image generation, we used the identical setup as depicted in Figure 2a.

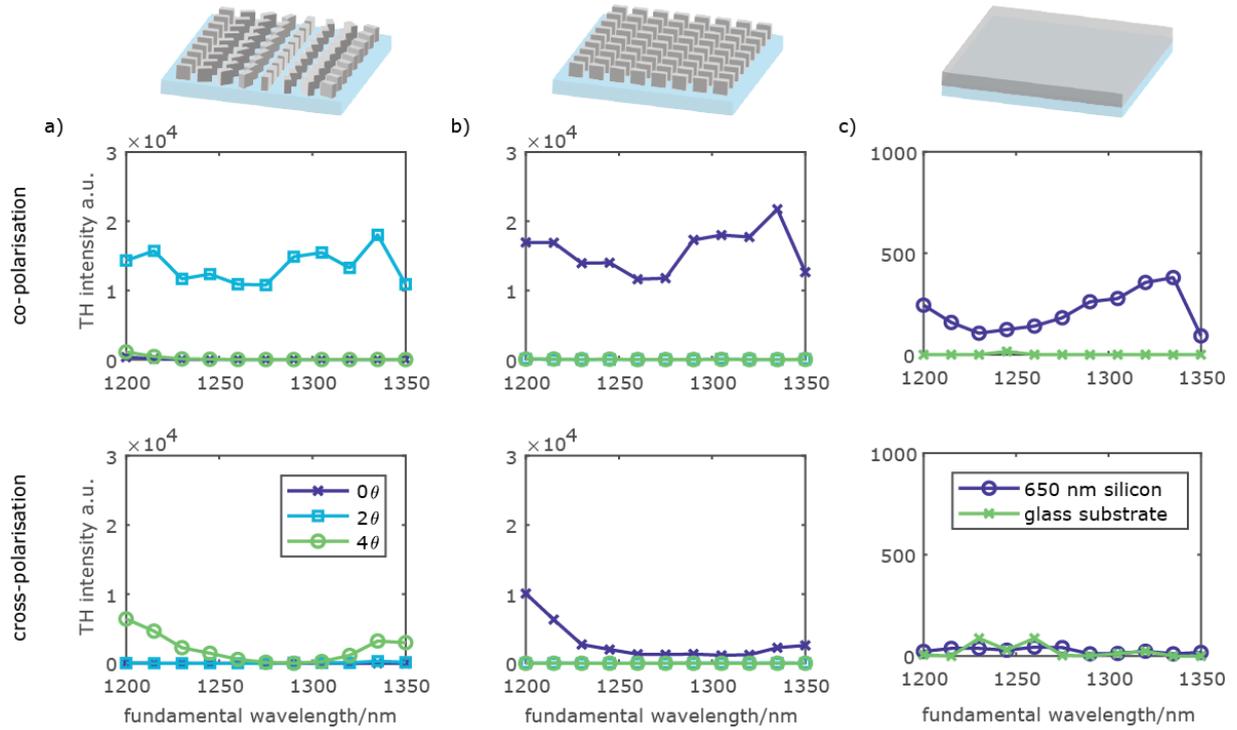

**Figure 3.** Measured THG intensity vs. wavelength for the different combinations of the input and output polarizations. **a)** Results for the gradient-phase metasurface showing a nearly constant THG signal. **b)** Results for the metasurface without phase gradient. Here, only the zeroth-order is analyzed since every nanofin has the same rotation and therefore, has the same phase. **c)** THG intensities of a 650-nm-thick amorphous silicon film and an ITO-coated glass substrate without silicon. The silicon film shows only a weak THG in the co-polarization states and a negligible signal in the cross-polarization. In comparison, the ITO-coated glass substrate shows a negligible THG signal in both polarization states.

The obtained results for the holograms at the THG wavelength are shown in Figure 4. The 'X'-hologram in Figure 4a is designed to be reconstructed in the co-polarization state within an angular range of ±20°. The cross-polarization states are not supposed to carry image information and therefore show no pattern in the Fourier space. Whereas in the case of the multiplexed hologram (Figure 4c) the holographic images reconstruct in both, co- and cross-polarization. Depending on the polarization combination, different pictures are reconstructed in the co-polarization (the sun) and in the cross-polarization (the cloud). The reconstructed holographic images are clearly observed in the Fourier space of the captured THG signal of the metasurface. The THG spot in the center of the images in Figure 4a arises from the residual THG signal that is not carrying the PB-phase. The measured holographic images exhibit some minor speckle pattern, which is typical for phase-only holograms. In addition, the imperfections and variations in the nanofins sizes from the sample fabrication and the illumination conditions can further enhance this effect. Note that by inverting the polarization state of the fundamental illumination the hologram will generate the conjugated image due to the change of the sign of $\sigma$ in the phase factor of the THG signal.

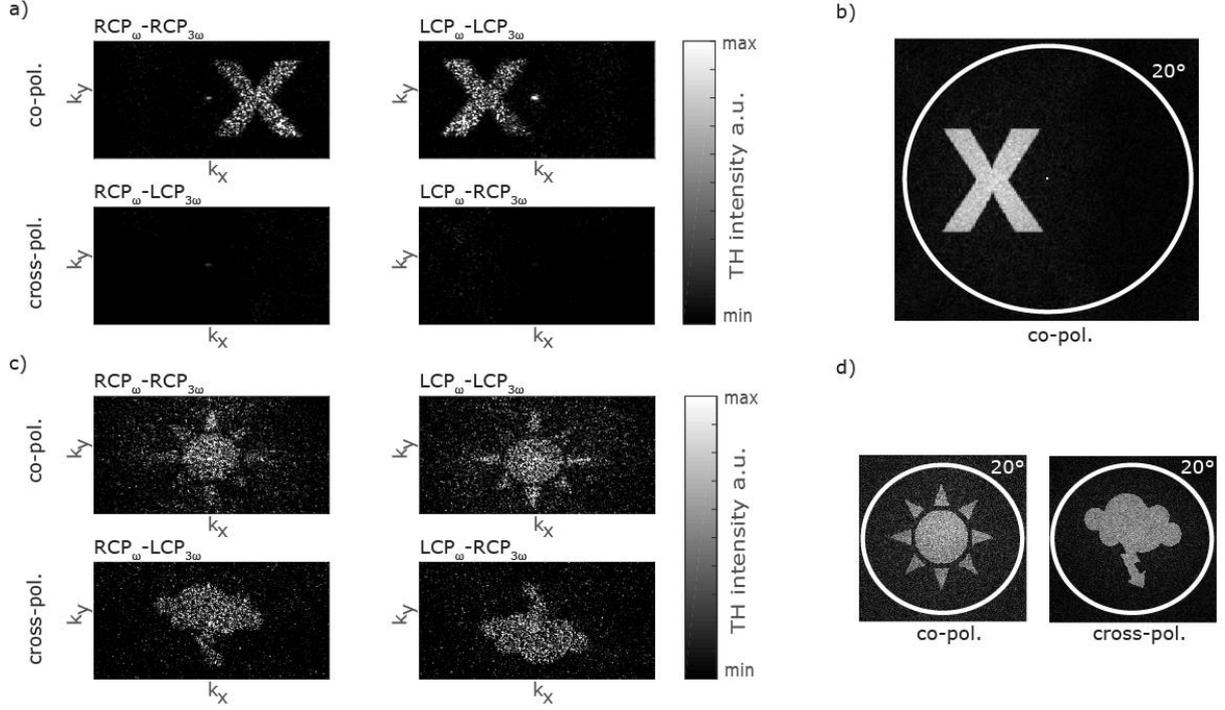

Figure 4. a) Measured holographic image of the letter 'X' for different combinations of the input and output polarizations. The image only appears in the co-polarization state of the THG signal while no image is observed in the cross-polarization state. b) The simulated holographic image of the letter 'X'. The image is designed to appear within an angle of 20° in the Fourier space. c) Measured multiplexed holographic image for different combinations of the input and output polarizations. d) The corresponding simulated holographic image of the sun and the cloud for both polarization states.

**Discussion**

The measurement results of the THG signals from the phase-gradient metasurface (Figure 2) demonstrate that a geometric phase (PB-phase) can be added during the nonlinear frequency conversion in dielectric nanofin structures. We encoded a linear phase gradient for the THG signal onto a nanostructured amorphous silicon metasurface in such a way that rotation of a single silicon nanofin induces a phase change of $\theta_{co} = 2\sigma\phi$ (co-pol.) and $\theta_{cross} = 4\sigma\phi$ (cross-pol.) in the THG signal, respectively, when the metasurface is illuminated with circularly polarized light. The encoded phase gradient leads to different nonlinear diffraction spots that appeared at the designed different angles. Since all nanofins have the same size and shape the phase in the nonlinear signal is controlled solely by the orientation (in plane rotation) of the nanofins. The Si metasurface shows negligible zeroth-order diffraction of the THG signal, which gives a high suppression of undesired THG signals that do not carry the desired phase information. Furthermore, our experiments demonstrate that two distinct phases appear in the co- and cross-polarized THG signals. Additionally, the phase information can be retrieved over a broad wavelength range with similar intensity. As shown in the Supplementary Material, the nanofins have a broad resonance in the range from 1200 nm to 1350 nm. However, the nonlinear response of the nanofins is featureless in the same range. These findings agree with the concept that the PB-phase depends on the rotation of the nanofins and only to a lesser extent on the resonance condition. This is different from previously published results on THG of Si nanopost metasurfaces. Here, only one phase factor was obtained by engineering the resonance and the working wavelength has to be fixed in advance. [25,24]

The comparison of the Si metasurface with an unstructured Si film shows that the patterning of the silicon leads to a new nonlinear behavior of the material. In an ideal isotropic film, THG for circularly polarized light is not possible within the dipole approximation and therefore no co- and cross-polarization THG signal should appear [29]. Nevertheless, the nanofins show THG in both the co- and cross-polarization states, which is a distinct behavior compared to the unstructured silicon thin film. Our findings are in agreement with previous works, stating that nanofins with C2 rotational symmetry generate third harmonic light in the co- and cross-polarization states [30,28,2]. The C2 rotational symmetry fulfills the THG selection rules for circularly polarized light, which explains why the THG is enhanced compared to the thin film. However, the measurements of the THG at the thin film show a weak THG in the co-polarization (Figure 3c) as we used hydrogenated amorphous silicon. Such plasma-enhanced chemical vapor (PECVD) deposited silicon results in defects in the film, which locally break the isotropy, leading to a weak nonzero THG.

To demonstrate that the PB-phase in dielectric metasurfaces can be utilized to tailor the phase of the THG, we encoded more complex phase pattern, resampling phase-only holograms, onto the metasurface (Figure 4). The holographic images can be reconstructed with high fidelity since the PB-phase approach allows for a continuous phase change in the full 0 to $2\pi$ range while keeping the THG amplitude unchanged. The measured holograms show only a weak zeroth-order THG spot in the centre of the image, which confirms the results obtained by the gradient-phase metasurface (Figure 2a) that nearly the entire THG signal carries the desired phase information. However, since the period of the nanofins is not subwavelength, the hologram appears at higher diffraction orders, again (for more details see Supplementary Material). Furthermore, the nonlinear hologram allows for polarization multiplexing as the two polarization states of the THG signal carry different phases. Hence, dynamically switching (either on/off or between two images) is possible by changing the observed polarization from co- to cross-polarization or vice versa. Similar multiplexing schemes have been already demonstrated for plasmonic metasurfaces [8].

Despite the relatively high packing density of the nanofins, which can result in strong crosstalk between the structures and the thickness of the nanofins exceeding the wavelength of the THG, which leads to phase matching issues. The experimental observations clearly demonstrate the possibility to tailor the nonlinear phase with identical dielectric nanostructures. Such tailoring is in analogy to the previously demonstrated nonlinear phase manipulation using the PB-phase with plasmonic nanostructures that show stronger field localization (less crosstalk) and smaller thicknesses of <50 nm (relaxing the phase-matching issue). However, the different electromagnetic modes inside the Si nanofins (which act as dielectric resonator) have different radiation patterns and field distributions, which not only complicates the phase control of the THG but also alters the conversion efficiency.[24,11,21,10] Therefore, the utilization of a geometric phase effect demonstrated here, can help to simplify the design of nonlinear dielectric metasurfaces.

## Summary

With our work, we experimentally demonstrate that a nonlinear phase manipulation by using the Pancharatnam-Berry phase concept in all-dielectric metasurfaces made of silicon nanofins is feasible. We encode phase information into the THG of a single silicon nano-resonator utilizing the rotation orientation and circularly polarized light. By building arrays of resonators, we demonstrate anomalous diffraction and complex beam shaping, like holograms, for the TH signals. The PB-phase approach for the nonlinear material polarization leads to a simplified design and eases the fabrication tolerances. Further, the high damage thresholds for dielectric metasurfaces make the design a suitable platform for complex beam shaping with nonlinear optical elements. For example, since both THG polarization states carry different phases and are orthogonal, the design can be used for nonlinear multiplexed holograms. Further applications include highly efficient nonlinear vector beam and orbital angular momentum generation, and nonlinear imaging.

## Acknowledgments

This project has received funding from the European Research Council (ERC) under the European Union's Horizon 2020 research and innovation programme (grant agreement No 724306), the Deutsche Forschungsgemeinschaft (DFG ZE953/11-1), and the National Natural Science Foundation of China (NSFC grant no. 61861136010).

## Supporting Information Available:

The Supplementary Material contains information on the fabrication of the metasurface, the phase grating design, the used algorithm to generate the phase profile of the holograms, details of the nonlinear measurements, the conversion efficiency, the position change of the diffraction order during a wavelength sweep, the input power dependence of the THG signal, the wavelength sweep for all polarization combinations, reappearance of the hologram at higher diffraction orders, and transmission spectra and field plots of the nano-resonator design.

**References**

1. Krasnok, A.; Tymchenko, M.; Alù, A. *Mater. Today* **2018,** *21* (1), 8–21, DOI:10.1016/j.mattod.2017.06.007.

2. Chen, S.; Li, G.; Zeuner, F.; Wong, W. H.; Pun, E. Y. B.; Zentgraf, T.; Cheah, K. W.; Zhang, S. *Optics Express* **2014,** *113* (3), 33901, DOI:10.1103/PhysRevLett.113.033901.

3. Kim, S.; Jin, J.; Kim, Y.-J.; Park, I.-Y.; Kim, Y.; Kim, S.-W. *Nature* **2008,** *453* (7196), 757–760, DOI:10.1038/nature07012.

4. Li, G.; Sartorello, G.; Chen, S.; Nicholls, L. H.; Li, K. F.; Zentgraf, T.; Zhang, S.; Zayats, A. V. *Laser Photonics Rev.* **2018,** *12* (6), 1800034, DOI:10.1002/lpor.201800034.

5. Schlickriede, C.; Waterman, N.; Reineke, B.; Georgi, P.; Li, G.; Zhang, S.; Zentgraf, T. *Advanced Materials (Deerfield Beach, Fla.)* **2018,** *30* (8), DOI:10.1002/adma.201703843.

6. Keren-Zur, S.; Avayu, O.; Michaeli, L.; Ellenbogen, T. *ACS Photonics* **2016,** *3* (1), 117–123, DOI:10.1021/acsphotonics.5b00528.

7. Chen, Y.; Yang, X.; Gao, J. *Advanced Optical Materials* **2018,** *6* (19), 1800646, DOI:10.1002/adom.201800646.

8. Ye, W.; Zeuner, F.; Li, X.; Reineke, B.; He, S.; Qiu, C.-W.; Liu, J.; Wang, Y.; Zhang, S.; Zentgraf, T. *Nature Communications* **2016,** *7*, 11930, DOI:10.1038/ncomms11930.

9. Li, G.; Zhang, S.; Zentgraf, T. *Nat. Rev. Mater.* **2017,** *2* (5), 17010, DOI:10.1038/natrevmats.2017.10.

10. Gili, V. F.; Carletti, L.; Locatelli, A.; Rocco, D.; Finazzi, M.; Ghirardini, L.; Favero, I.; Gomez, C.; Lemaître, A.; Celebrano, M.; Angelis, C. de; Leo, G. *Optics Express* **2016,** *24* (14), 15965–15971, DOI:10.1364/OE.24.015965.

11. Carletti, L.; Koshelev, K.; Angelis, C. de; Kivshar, Y. *Physical Review Letters* **2018,** *121* (3), 33903, DOI:10.1103/PhysRevLett.121.033903.

12. Golla, C.; Weber, N.; Meier, C. *J. Appl. Phys.* **2019,** *125* (7), 73103, DOI:10.1063/1.5082720.

13. Shcherbakov, M. R.; Neshev, D. N.; Hopkins, B.; Shorokhov, A. S.; Staude, I.; Melik-Gaykazyan, E. V.; Decker, M.; Ezhov, A. A.; Miroshnichenko, A. E.; Brener, I.; Fedyanin, A. A.; Kivshar, Y. S. *Nano Letters* **2014,** *14* (11), 6488–6492, DOI:10.1021/nl503029j.

14. Xu, L.; Rahmani, M.; Zangeneh Kamali, K.; Lamprianidis, A.; Ghirardini, L.; Sautter, J.; Camacho-Morales, R.; Chen, H.; Parry, M.; Staude, I.; Zhang, G.; Neshev, D.; Miroshnichenko, A. E. *Light Sci. Appl.* **2018,** *7* (1), 917, DOI:10.1038/s41377-018-0051-8.

15. Grinblat, G.; Li, Y.; Nielsen, M. P.; Oulton, R. F.; Maier, S. A. *Nano Letters* **2016,** *16* (7), 4635–4640, DOI:10.1021/acs.nanolett.6b01958.

16. Campione, S.; Liu, S.; Basilio, L. I.; Warne, L. K.; Langston, W. L.; Luk, T. S.; Wendt, J. R.; Reno, J. L.; Keeler, G. A.; Brener, I.; Sinclair, M. B. *ACS Photonics* **2016,** *3* (12), 2362–2367, DOI:10.1021/acsphotonics.6b00556.

17. Yang, Y.; Wang, W.; Boulesbaa, A.; Kravchenko, I. I.; Briggs, D. P.; Puretzky, A.; Geohegan, D.; Valentine, J. *Nano Letters* **2015,** *15* (11), 7388–7393, DOI:10.1021/acs.nanolett.5b02802.

18. Sautter, J. D.; Xu, L.; Miroshnichenko, A. E.; Lysevych, M.; Volkovskaya, I.; Smirnova, D. A.; Camacho-Morales, R.; Zangeneh Kamali, K.; Karouta, F.; Vora, K.; Tan, H. H.; Kauranen, M.; Staude, I.; Jagadish, C.; Neshev, D. N.; Rahmani, M. *Nano Letters* **2019,** *19* (6), 3905–3911, DOI:10.1021/acs.nanolett.9b01112.

19. Liu, S.; Vabishchevich, P. P.; Vaskin, A.; Reno, J. L.; Keeler, G. A.; Sinclair, M. B.; Staude, I.; Brener, I. *Nature Communications [Online]*, *9* (1), 2507. https://www.nature.com/articles/s41467-018-04944-9.pdf.

20. Camacho-Morales, R.; Rahmani, M.; Kruk, S.; Wang, L.; Xu, L.; Smirnova, D. A.; Solntsev, A. S.; Miroshnichenko, A.; Tan, H. H.; Karouta, F.; Naureen, S.; Vora, K.; Carletti, L.; Angelis, C. de; Jagadish, C.; Kivshar, Y. S.; Neshev, D. N. *Nano Letters* **2016,** *16* (11), 7191–7197, DOI:10.1021/acs.nanolett.6b03525.

21. Ghirardini, L.; Marino, G.; Gili, V. F.; Favero, I.; Rocco, D.; Carletti, L.; Locatelli, A.; Angelis, C. de; Finazzi, M.; Celebrano, M.; Neshev, D. N.; Leo, G. *Nano Letters* **2018,** *18* (11), 6750–6755, DOI:10.1021/acs.nanolett.8b02432.

22. Grinblat, G.; Li, Y.; Nielsen, M. P.; Oulton, R. F.; Maier, S. A. *ACS Photonics* **2017,** *4* (9), 2144–2149, DOI:10.1021/acsphotonics.7b00631.

23. Kauranen, M.; Zayats, A. V. *Science Advances* **2012,** *6* (11), 737–748, DOI:10.1038/nphoton.2012.244.

24. Wang, L.; Kruk, S.; Koshelev, K.; Kravchenko, I.; Luther-Davies, B.; Kivshar, Y. *Nano Letters* **2018,** *18* (6), 3978–3984, DOI:10.1021/acs.nanolett.8b01460.

25. Gao, Y.; Fan, Y.; Wang, Y.; Yang, W.; Song, Q.; Xiao, S. *Nano Letters* **2018,** **DOI:**10.1021/acs.nanolett.8b04311.

26. Guo, Q.; Schlickriede, C.; Wang, D.; Liu, H.; Xiang, Y.; Zentgraf, T.; Zhang, S. *Optics Express* **2017,** *25* (13), 14300–14307, DOI:10.1364/OE.25.014300.

27. Bar-David, J.; Levy, U. *Nano Letters* **2019,** *19* (2), 1044–1051, DOI:10.1021/acs.nanolett.8b04342.

28. Li, G.; Chen, S.; Pholchai, N.; Reineke, B.; Wong, P. W. H.; Pun, E. Y. B.; Cheah, K. W.; Zentgraf, T.; Zhang, S. *Nature Materials* **2015,** *14* (6), 607–612, DOI:10.1038/nmat4267.

29. Tang, C. L.; Rabin, H. *Phys. Rev. B* **1971,** *3* (12), 4025–4034, DOI:10.1103/PhysRevB.3.4025.

30. Bhagavantam, S.; Chandrasekhar, P. *Proc. Indian Acad. Sci.* **1972,** *76* (1), 13–20, DOI:10.1007/BF03048332.